\documentclass{article}

\usepackage{fullpage}
\usepackage[fleqn,reqno]{amsmath}
\usepackage{amsfonts, amssymb, stmaryrd, bm}
\usepackage{graphicx}
\usepackage{color}
\usepackage{subcaption}
\usepackage[hidelinks]{hyperref}
\usepackage{algorithmic}
\usepackage{algorithm}
\usepackage{authblk}

\newcommand{\x}{\mathbf{x}}
\newcommand{\R}{\mathbb{R}}
\newcommand{\n}{\mathbf{n}}
\newcommand{\crgb}{\mathbf{c}}

\title{A Survey of Methods for Converting Unstructured Data to CSG Models}
\author[1]{Pierre-Alain Fayolle}
\author[2]{Markus Friedrich}
\affil[1]{University of Aizu, Japan}
\affil[2]{Hochschule M{\"u}nchen, Germany}
\date{}

\begin{document}
\maketitle

{\bf Abstract:} The goal of this document is to survey existing methods for recovering CSG representations from unstructured data such as 3D point-clouds or polygon meshes. We review and discuss related topics such as the segmentation and fitting of the input data. We cover techniques from solid modeling and CAD for polyhedron to CSG and B-rep to CSG conversion. We look at approaches coming from program synthesis, evolutionary techniques (such as genetic programming or genetic algorithm), and deep learning methods. Finally, we conclude with a discussion of techniques for the generation of computer programs representing solids (not just CSG models) and higher-level representations (such as, for example, the ones based on sketch and extrusion or feature based operations).

\section{Introduction}
\begin{figure}[!h!t!bp]
\centering
\begin{subfigure}[h]{0.3\textwidth}
  \includegraphics[width=\linewidth]{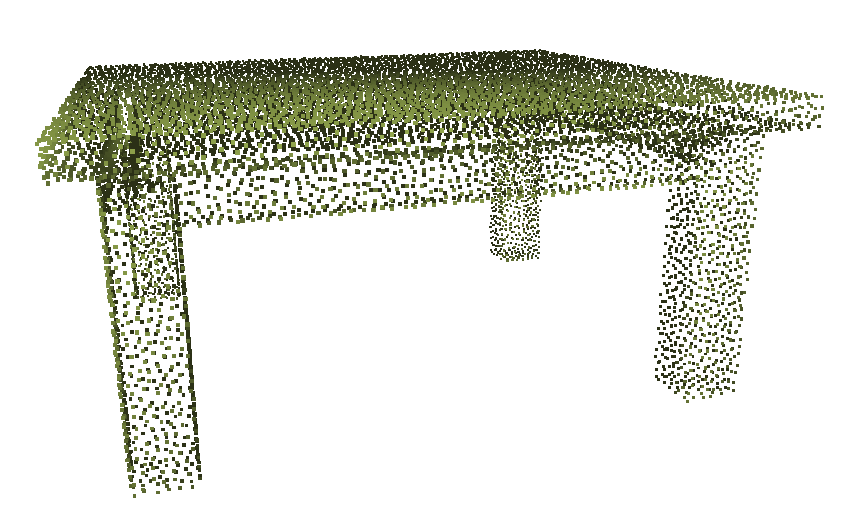}
\end{subfigure}
\begin{subfigure}[h]{0.3\textwidth}
  \includegraphics[width=\linewidth]{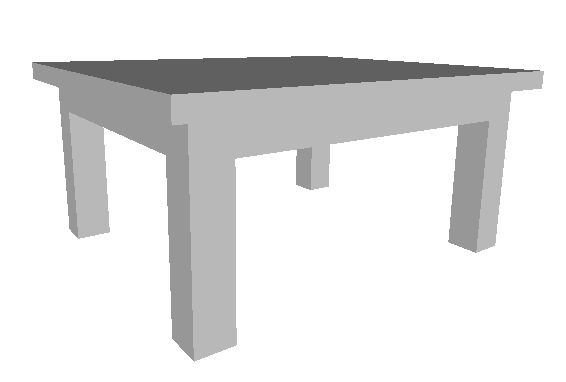}
\end{subfigure}
\begin{subfigure}[h]{0.3\textwidth}
  \includegraphics[width=\linewidth]{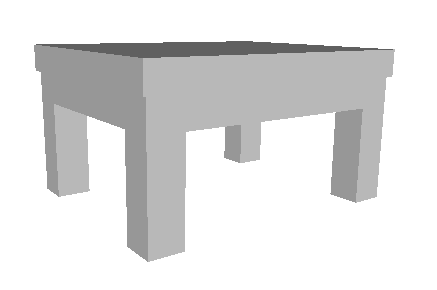}
\end{subfigure}
\caption{Left: 3D point-cloud corresponding to a table. Middle: A CSG model for the table. Right: An edited version of the model, where the width, depth and height of the table-top were modified.}\label{fig:table}
\end{figure}

There is an increasing availability of devices for acquiring 3D data sets: Laser scanners, LIDAR, RGBD cameras, or even regular RGB cameras with the help of photogrammetry \cite{Hartley2003}, resulting in the availability of large collections of 3D data sets. A polygon mesh can be obtained from a given input 3D point-cloud by using a surface reconstruction algorithm. It often involves building an implicit surface via a collection of splines or polynomials as in \cite{carr2001,ohtake2003}, or as samples on a (structured or unstructured) grid as in \cite{Kazhdan2006}. Recently, the focus has been on using techniques from deep learning, with the implicit surface defined as the zero level-set of a multi-layer perceptron (MLP). See, for example, the approach described in \cite{sitzmann2020}. The main problem with these methods is that the output that they produce (polygon mesh, weighted sum of polynomials/splines, MLP) is not an editable model. Ideally, a user should have control over the recovered model, and should be able to edit it, store it, fabricate it (using, for example, a 3D printer). An example of a 3D point-cloud for a table, obtained by photogrammetry, a CSG model of the object and an edited version of the CSG model is shown in Fig.\,\ref{fig:table}. Of course, a CSG-based recovery approach does not necessarily apply to all types of 3D objects, and some of the solids with a freeform shape are better handled with the former approaches. 

\subsection{Objectives}
The main objective of this survey is to cover the existing approaches for recovering a CSG model from an unstructured data source. We consider 3D point-clouds and polygon (triangle or quadrilateral) meshes or soups as the prototypical input data, though one could of course consider as well a collection of images that can be converted to a 3D point-cloud via photogrammetry. Since they are related problems, we will also consider the existing algorithms for converting a polyhedron to a CSG model (involving half-planes only) and the existing techniques for converting a B-rep (with possibly curved faces) to a CSG model. We will also discuss techniques for the generation of computer programs representing solids (procedural shape/solid programs) and higher level representations (such as, for example, sketch and extrusion). 

\subsection{Organization}
This manuscript is organized as follows: We start by providing in Section~\ref{sec:background} some details on tools and techniques used elsewhere in the paper. This includes information on the input data and the primitives used (Section~\ref{sec:input}) and on the CSG representation (Section~\ref{sec:csg}). In Section~\ref{sec:CSGinCAD}, we deal with methods for converting polyhedra and B-rep models to CSG expressions. Section~\ref{sec:methods} deals with techniques for extracting (or recovering) a CSG model from unstructured data. We describe techniques for shape decomposition and primitives fitting (Section~\ref{sec:fitting}) and approaches for recovering general CSG expressions (Section~\ref{sec:generation}).
Finally, we propose a discussion of techniques for generating higher-level representations and shape procedures and conclude in Section~\ref{sec:conclusion}.

\section{Preliminaries}\label{sec:background}
In this section we provide background material on the terms and techniques used in the rest of this document. We describe the types of data that we are dealing with, as well as a description of the CSG representation. 

\subsection{Input data}\label{sec:input}
We deal in this work with unstructured 3D data. This includes point-clouds and polygon meshes or soups, with the most common cases consisting of triangles and quadrilaterals. 

\begin{figure}[!h!t!bp]
\centering
\begin{subfigure}[h]{0.3\textwidth}
  \includegraphics[width=\linewidth]{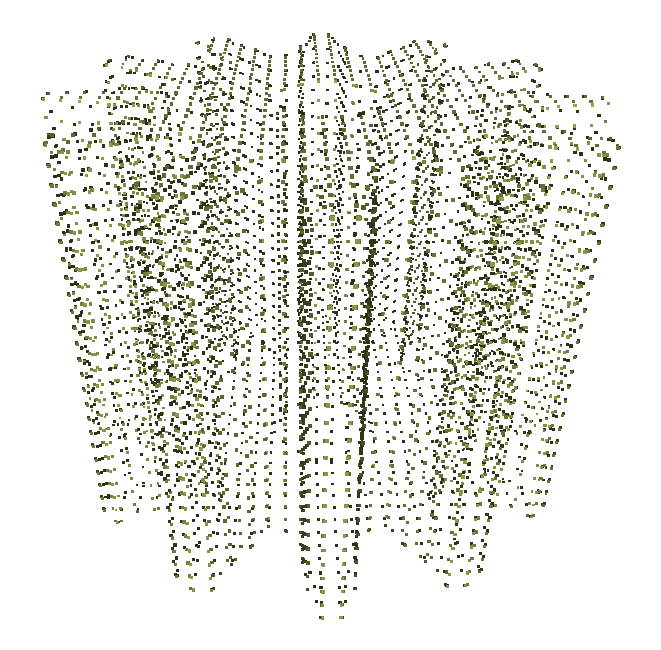}
  \caption{A 3D point-cloud.}
\end{subfigure}
\begin{subfigure}[h]{0.3\textwidth}
  \includegraphics[width=\linewidth]{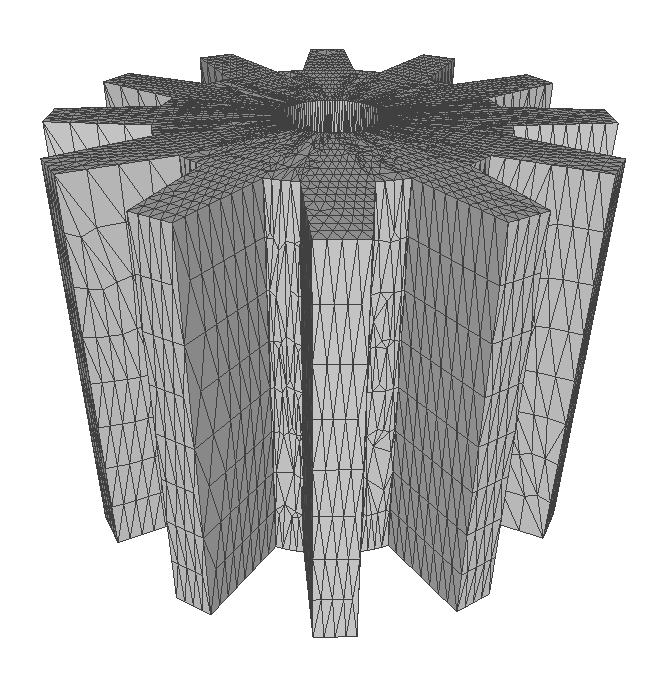}
  \caption{A triangle mesh.}
\end{subfigure}
\begin{subfigure}[h]{0.3\textwidth}
  \includegraphics[width=\linewidth]{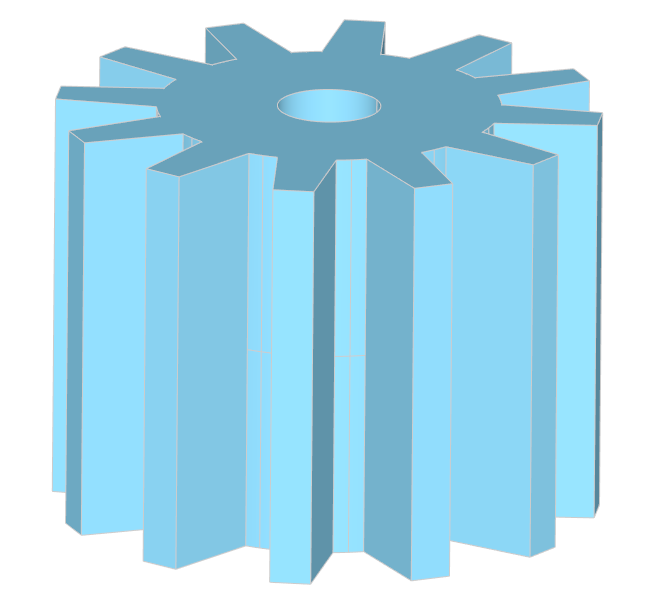}
  \caption{A B-rep model.}
\end{subfigure}
\caption{Examples of inputs and data types.}\label{fig:input}
\end{figure}

\subsubsection{Point-clouds}
A 3D point-cloud is a collection of points $\{\x_1, \x_2, \ldots \x_n\}$, where $\x_i \in \R^3$ is a point coordinates, for $i=1, \ldots, n$. Often, the points come with additional attributes such as normal vectors, $\n_i \in \R^3$, or color attributes, $\crgb_i \in [0,1]^3$. The normal vectors are useful for geometric problems, since they allow to orientate an object. We will omit the color attributes, since we deal in this work with a geometric problem. If the normal vectors are not captured by the sensor, they can be approximated afterward. The classical method for estimating the normal vectors and their orientation was introduced by Hoppe et al. in \cite{Hoppe1992}. Recently, methods based on deep learning have been investigated, such as, for example, PCPNET in \cite{Guerrero2018}. An example of a 3D point-cloud for a mechanical part is shown in Fig.\,\ref{fig:input}(a). 

\subsubsection{Polygon meshes and boundary representation}
Polygon meshes are ubiquitous in computer graphics. They consist in a collection of polygons specified by their end-points coordinates. Most commonly, collections of triangles and/or quadrilaterals are used. Typically, a triangle mesh will be defined as a list of vertex coordinates $V=\{\x_1, \x_2, \ldots, \x_n\}$ and a list of triangles $T=\{\ldots,(i,j,k),\ldots\}$, where $i,j,k$ are indices in the list of vertex coordinates $V$. An example of a triangle mesh is shown in Fig.\,\ref{fig:input}(b). 

In engineering, the Boundary representation (B-rep) is the preferred representation for a solid. A B-rep model consists in a collection of topological entities: vertices, edges, faces, their connection, and their geometric realisation (point coordinates, curve definition, surface patch definition). An example of B-rep model is shown in Fig.\,\ref{fig:input}(c). B-rep is a much richer representation than a polygon mesh based representation. In this work, we are interested in the problem of recovering a model from unstructured data, and assume that our input is a 3D point-cloud or an unstructured collection of triangles (a so-called polygon soup). 

\subsubsection{Data set collections}\label{sec:dataset}
Over the years, collections of structured and unstructured point-clouds, triangle and polygon meshes have been accumulated, curated and organized for purposes such as benchmarking algorithms or training data-driven algorithms. Thingi10K \cite{zhou2016thingi10k} is a collection of triangle meshes, designed for benchmarking meshing algorithms. It consists in a collection of models from Thingiverse \cite{thingiverse}. ShapeNet \cite{chang2015shapenet,wu20153d} is targeting machine learning algorithms and computer vision applications, and is limited to a few families (or categories) of objects. The ABC data set was introduced in \cite{Koch2019} as a large collection of CAD models. The models are available in different formats such as: polygon mesh and B-rep. This data set was developed for the purpose of training data driven algorithms for geometric deep learning methods and applications. The objects are sampled from OnShape's public models \cite{onshape}. Similarly, the Fusion 360 data set \cite{Willis2021} is a collection of CAD models / programs corresponding to sequences of sketches/extrusions. The objects are sampled from models designed with Fusion 360 \cite{fusion360}. This data set was designed with the aim to train data driven algorithms to learn CAD models as short programs, i.e., programs consisting of sketches and extrusions. The data set assembled for DeepCAD \cite{wu2021deepcad} is based on ABC's data set, but focuses on a representation based on sketch curves and extrusions, similarly to the data set introduced in \cite{Willis2021}. Finally, Fit4CAD \cite{Romanengo2021fit4cad} introduces a benchmark for evaluating and comparing methods for fitting simple geometric primitives in point clouds corresponding to CAD objects. The collection of large data sets has two main objectives: 1) To give a common data set for benchmarking and comparing different algorithms and 2) to provide a sufficient amount of data for training machine learning algorithms. The second goal seems to be the predominant reason for collecting 3D data nowadays, and the data sets made recently available, such as \cite{Koch2019,Willis2021,wu2021deepcad}, have sparked the creation of several deep learning based approaches. 

\subsection{Primitives}\label{sec:primitives}
Geometric primitives can serve as basic building blocks for representing CAD geometries. Commonly used primitives include spheres, cylinders, cones, torii, planes and other so-called quadrics whose surfaces can be described as the zero level-set of a second-degree polynomial. More complex primitives can be used depending on the selected representation model, and the system being used. For the task of recovering a model from an unstructured data set, the complexity of the supported primitives is restricted by the segmentation and fitting approach used, see Section~\ref{sec:fitting}. In general, nothing prevents us to use any SDF (Signed Distance Function, a function $f(\x)$ that returns the signed distance from $\x$ to a given surface $\partial S$) as a primitive, as long as it can be identified and fitted in the input data set. 

\paragraph{Halfspaces}
The term \emph{halfspace} is used to denote a set $p = \{\x \in \R^3: f(\x) \le 0\}$, for a function $f: \R^3 \rightarrow \R$. The halfspace complement $\bar{p}$ is given by $\bar{p} = \{\x \in \R^3: f(\x) > 0\}$. When $f$ is a polynomial, the term \emph{polynomial halfspace} is used. The case where $f$ corresponds to the distance from $\x \in \R^3$ to the surface $\partial S$ of a solid $S$ is what is referred to an SDF. In general, there is not a universal convention on the sign of $f$ and on whether it corresponds to the interior or the exterior of the solid $S$. 

\subsection{CSG representation}\label{sec:csg}
A widely-known and intuitive representation of solids combines Boolean set-operations and geometric primitives in a tree-like structure. It is called a CSG (Constructive Solid Geometry) representation, and the corresponding tree structure is called a CSG tree. 

The set-operations commonly used in CSG modeling packages are typically the union, intersection and difference. In order to guarantee that operations on solids always yield solids, so-called regularized set-operations are used \cite{Requicha80}. The expressiveness of these so-called Constructive Solid Geometry (CSG) trees is highly dependent on the set of supported primitives. 

Given a solid with point-set $S$, a primitive-set $P$ and a set of binary operations $R$, the solid is fully described by a CSG tree expression $\Phi$ iff $|\Phi(P)|=S$, where $|\cdot|$ describes the point-set that results from the CSG expression $\Phi$ and primitives $P$. The formalism that we use here is derived from \cite{Shapiro1991cad}.

\subsubsection{Counting CSG trees and complexity}
Let $|P|$ be the number of primitives, i.e., the cardinal of the set of primitives $P$. For a given number $n$ of inner nodes, the number of binary trees is 
\[
C_n = \frac{1}{n+1}\binom{2n}{n},
\]
the so-called Catalan numbers. 
We consider for a CSG tree a binary tree with internal nodes tagged by the union or intersection operation, and each leaf tagged by either a primitive or its complement. The number of CSG trees for a given number of inner node $n$ is 
\[
\frac{1}{n+1}\binom{2n}{n} \cdot 2^n \cdot (2|P|)^{n+1}, 
\]
obtained by counting the number of binary trees and the possibilities for the inner nodes and the leaves. 
The number of possible trees with a number of inner nodes between $n_{\min}$ and $n_{\max}$ is given by (see also \cite{friedrich2022full})
\begin{equation} \label{eq:csgsearchspace}
	\sum_{n=n_{\min}}^{n_{\max}} (2|P|)^{n+1} \cdot 2^n \cdot\frac{1}{n+1} \binom{2n}{n}.
\end{equation}
Each primitive (or its complement) should appear at least once in the expression (tree), so we can assume $n_{\min}=|P|-1$. However, $n_{\max}$ is not known in general and must be estimated empirically. Equation \ref{eq:csgsearchspace} corresponds to the size of the search space of the CSG generation problem. 

There exist several different sub- and super-sets of CSG expressions with binary operations leading to different complexities (size for the search space). 

\subsubsection{DNF Expressions (DE, sub-set)}
Restricting the set of expressions to those in Disjunctive Normal Form (DNF), we obtain
\begin{equation}\label{eq:canonicalcuts}
\Phi(P) = \bigcup_{k=1}^{2^{|P|}-1} \epsilon_k \left( g_1 \cap^* g_2 \dots \cap^* g_{|P|} \right), \qquad g_i \in \{p_i, \setminus^{*}p_i\}, 	
\end{equation}
where $\epsilon_k$ is $1$, if the point-set of the $k^\text{th}$ fundamental product $(g_1 \cap^* g_2 \dots \cap^* g_{|P|})$ is inside $S$ and $0$ otherwise. This reduces the search space substantially to $O(2^{|P|})$. However, the optimality of the expression size is not guaranteed anymore. 
\\
Note that a known result in Boolean logic states that a Boolean function can always be written in full disjunctive normal form. Thus, in theory we can always search for a CSG expression in the form (\ref{eq:canonicalcuts}), and it gives an upper bound for the complexity. 

\subsubsection{Union Expressions (UE, sub-set)}
Under the assumption that all primitives are part of the final solid, the resulting CSG expression is simply the union of all primitives: 
\begin{equation}\label{eq:canonicalcutssimp}
\Phi(P) = \bigcup_{k=1}^{|P|} p_i 	
\end{equation}
Since the union operation is symmetric, the tree structure does not matter. This is the assumption made by most of the shape decomposition methods (see Section~\ref{sec:deep_fitting}).

\subsubsection{Tree Expressions for Decomposable Solids (DTE, sub-set)}
A solid is said to be {\em decomposable} if its primitives are all located either fully inside or fully outside of the solid (so-called dominant primitives in \cite{Shapiro1991cad}).
These primitives can be removed (factored) from the solid in a recursively executed decomposition process: 
\begin{equation}
\label{eq:dec}
S^i = ((...( S^{i+1}\oplus |d^i_1| ) \oplus ...) \oplus |d^i_{n-1}|) \oplus |d^i_n|,
\end{equation}
where $S^i$ is the point set of the remaining solid and $D^i=\{d^i_1,...,d^i_n\}$ is the sequence of dominant primitives for recursion level $i$ with $|d|$ denoting the point set induced by primitive $d$. $\oplus$ is either the set difference operation (if the following dominant primitive is fully outside $S^i$) or the set union operation otherwise. In the case of a fully decomposable solid, the search space of all possible CSG expressions shrinks down to $O(|P|^2)$. Furthermore, it is guaranteed that the resulting expression is size-optimal. However, not all solids are decomposable in this way. 

\subsubsection{m-ary Tree Expressions (MTE, super-set)}
If $m$-ary operations such as, for example, the unary complement (also called negation) should be supported, the tree structure must allow different numbers of child nodes which makes it substantially more difficult to estimate the search space size. 

If we consider as operations: union, intersection, difference (binary) and the complement (unary), we have a tree where each node can have 0, 1 or 2 descendants. Such a tree is called a unary-binary tree. 

The number of unary-binary trees with $n$ edges (or equivalently with $n+1$ nodes) is given by the Motzkin numbers 
\[
M_n = \frac{1}{n}\sum_{k=0}^{n-1} \binom{n}{k} \binom{k}{n-1-k}. 
\]

However, if we want to count the number of unary-binary trees with $n$ inner nodes, it is given by the $n$-th large Schroeder number $S_n$. Unlike the Motzkin number $M_n$, there is no closed form expression for the large Schroeder number, and it is given by the following relation
\[
(n+1) S_n = 3 (2n-1) S_{n-1} - (n-2) S_{n-2}. 
\]
If we denote by $u$ the number of unary operators, $b$ the number of binary operators and $L$ the number of leaves ($L = |P|$ or $L=2|P|$, where $|P|$ is the number of primitives), the number of expressions for $n$ inner nodes is given by the following relation 
\[
(n+1) T_n = (u + 2bL)(2n-1) T_{n-1} - u (n-2) T_{n-2}. 
\]
While there are no closed form expressions for $S_n$ and $T_n$, it is possible to compute asymptotic estimation \cite{flajoletSedgewick}. Similarly to the binary case, the search space is obtained by considering all possible number of inner nodes: $\sum_n T_n$.

\subsubsection{Graph Expressions (GE, super-set)}
If the space of tree-based expressions is extended to graph-based structures, well-known programming constructs like loops are possible. This widens the search space of possible expressions to that of programs written in higher-level CAD-specific programming languages and methods from program synthesis come into play. 

Here again, we can use asymptotics to get the following estimation of the number of (unlabelled) graphs with $n$ inner nodes 
\[
G_n ~ \frac{1}{n!} 2^{\binom{n}{2}}, 
\]
where $G'_n=2^{\binom{n}{2}}$ is the number of labelled graphs, and we have asymptotically that $G_n \sim 1/n! G'_n$. 

\subsection{CSG and SDF}
\label{sec:csgsdf}
Given a CSG expression, we can form an SDF (strictly speaking, an approximation only) by replacing the primitives with their SDF representations and formulating the Boolean operations as min/max-functions:
\begin{align}
	|\Phi| \cap^{*} |\Psi| &:= \max(f_{\Phi}, f_{\Psi})\\
	|\Phi| \cup^{*} |\Psi| &:= \min(f_{\Phi}, f_{\Psi}) \\
	\setminus^{*} |\Phi| &:= -f_{\Phi} \\
	|\Phi| -^{*} |\Psi| &:= \max(f_{\Phi}, -f_{\Psi}),
\end{align}
Note that the use of min and max would be exchanged if a different orientation of the primitives is selected. The expressions above assume that the solid $|\Phi|$ corresponds to the point-set $\{\x: f_{\Phi}(\x) \le 0\}$. 
An alternative formulation to the min/max-functions are the so-called R-Functions \cite{shapiro1991theory}:
\begin{align}
|\Phi| \cap^{*} |\Psi| &:= \frac{1}{2} ( f_{\Phi} + f_{\Psi} + \sqrt{(F_{\Phi} - F_{\Psi})^2 + \alpha}) \\
|\Phi| \cup^{*} |\Psi| &:=\frac{1}{2} ( f_{\Phi} + f_{\Psi} - \sqrt{(F_{\Phi} - f_{\Psi})^2 + \alpha})  \\
\setminus^{*} |\Phi| &:= -f_{\Phi} \\
|\Phi| -^{*} |\Psi| &:= \frac{1}{2} ( f_{\Phi} - f_{\Psi} - \sqrt{(f_{\Phi} - f_{\Psi})^2 + \alpha}),
\end{align}
with $\alpha \in \mathbb{R}$. Setting $\alpha=0$ gives the min/max-functions.

\section{CSG conversion methods in computational geometry and CAD}
\label{sec:CSGinCAD}

\subsection{Polyhedron to CSG} 
Given a polyhedron with $p$ faces, Paterson and Yao gave in \cite{Paterson1990} an algorithm with $O(p^3)$ time complexity to generate a CSG expression with size $O(p^2)$, but only for a restricted class of polyhedra. In 2D, a polygon can be converted to a CSG expression with $O(p)$ terms, this is called a Peterson-style formula \cite{Peterson1986}. Dobkin et al. gave a $O(p \log p)$ algorithm to compute a Peterson-style CSG formula for a given 2D polygon in \cite{Dobkin1988}; they also proved that in 3D not all polyhedra have a Peterson-style formula. Dey proved a lower bound of $O(p^2)$ on the size \cite{dey1991}. The construction of the CSG expression is based on a convex decomposition algorithm. For each detected feature that causes a non-convexity in a polyhedron, an additional plane is added to remove the non-convexity. Dey calls \emph{notches} the features causing non-convexity, and he calls \emph{notch planes} these additional planes. The resulting CSG expression is given by the union of the convex cells obtained from the convex decomposition algorithm. It is thus in DE form (\ref{eq:canonicalcuts}). Recently, Rossignac proposed to use an equivalence Boolean operation, and called the corresponding representation MAS (for Match Aggregate of Sectors) \cite{Rossignac_2022}. 

\subsection{B-rep to CSG}
Works on converting polyhedra to CSG led to the more difficult problem of B-rep to CSG conversion. The pioneering work was done by Shapiro and Vossler, first in 2D in  \cite{Shapiro1991jmd,Shapiro1991cad,Shapiro2001}, then extended to the 3D case in \cite{Shapiro1991cad,Shapiro1993}. Conversion from B-rep to BSP, and BSP to CSG was done by Buchele in \cite{Buchele1999}, and extended and improved in \cite{Buchele2004}. Some of the ideas and concepts introduced by Shapiro and Vossler in their works are fundamental and used as the basis for other works on point-cloud to CSG recovery. 

\subsubsection{Natural halfspaces, separating halfspaces and describability}
\label{sec:describability}
\begin{figure}[!h!t!bp]
\centering
\includegraphics[width=0.9\textwidth]{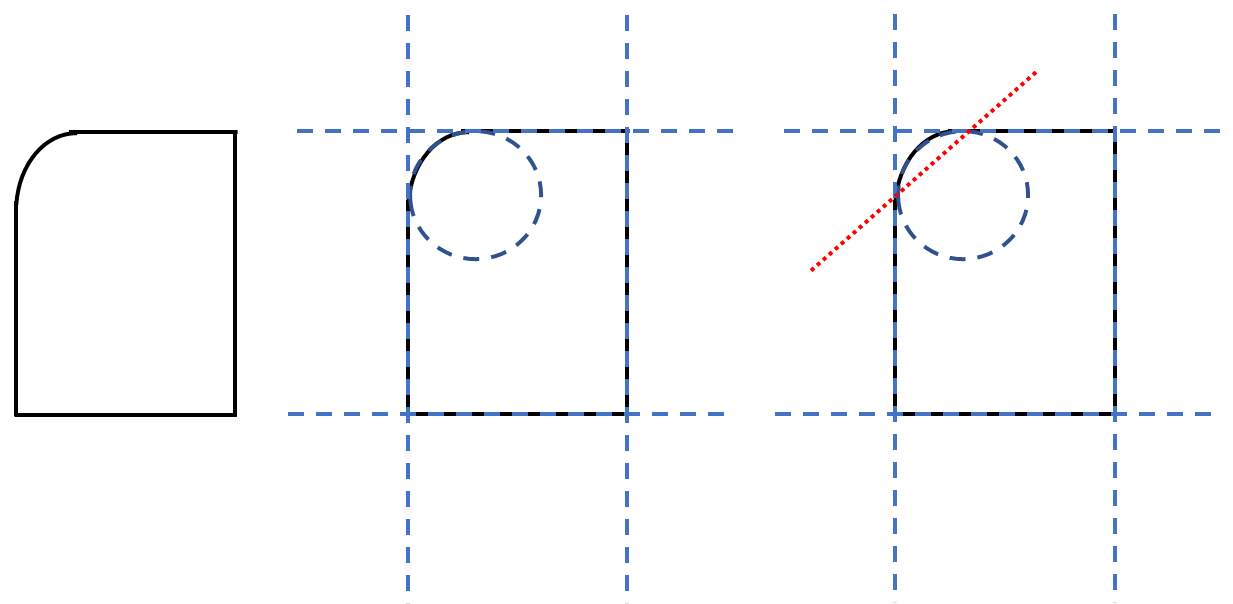}
\caption{Illustration of natural and separating halfspaces. Left: A 2D solid. Middle: The natural halfspaces are shown in dashed lines (blue color). Right: The separating halfspaces are shown in dotted lines (red color).}\label{fig:halfspaces}
\end{figure}
\begin{figure}[htb]
  \begin{center}
     \includegraphics[width=0.48\textwidth]{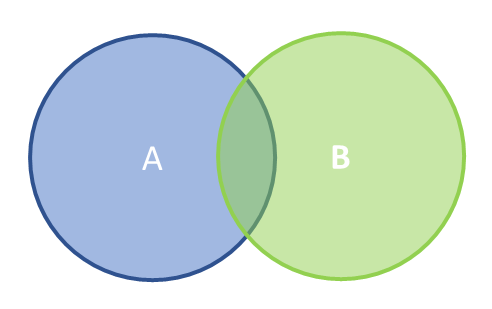}
   \end{center}
   \caption{Illustration of fundamental products. Given the sets $A$ and $B$, the fundamental products are: $A \cap B$, $\overline{A} \cap B$, $A \cap \overline{B}$ and $\overline{A} \cap \overline{B}$.}\label{fig:canonical}
\end{figure}

{\em Natural halfspaces} are induced by the natural surfaces of a solid, i.e. it is the minimal set of surfaces describing the boundary of a solid. 
A {\em separating halfspace} is a a halfspace that is necessary in the CSG representation of a solid but is not a natural halfspace of the solid. Figure \ref{fig:halfspaces} illustrate the concepts of natural and separating halfspaces. 
A set of $n$ halfspaces partitions $\mathbb{R}^3$ in $2^n$ subsets. We call each of these sets a {\em fundamental product} (or canonical cell or canonical intersection term or minterm). See Fig.\,\ref{fig:canonical} for an illustration of the concept of fundamental product. 
Shapiro and Vossler introduced also the concept of {\em describability}. Given a set of halfspaces, a solid is describable by this set if and only if for each fundamental product, all points in the fundamental product have the same point membership classification with respect to the solid.

\subsubsection{CSG generation and minimization}
\label{sec:brep2csg}
We are now ready to provide a first algorithm for generating a CSG expression from a given B-rep: A solid is defined as the union of the fundamental products that are contained entirely inside the solid. This idea is formalized in Algorithm \ref{brep2csg1}. 
\begin{algorithm}
\begin{algorithmic}
\caption{B-rep to CSG conversion using fundamental products}\label{brep2csg1}
\STATE Construct a set of halfspaces that is sufficient for a CSG representation of a solid. 
\STATE Find all fundamental products completely inside the solid. 
\STATE Take the union of all these fundamental products. 
\end{algorithmic}
\end{algorithm}

This algorithm simply expresses the fact in Boolean logic that any Boolean function can be expressed in full disjunctive normal form. It is also directly related to the DE representation (\ref{eq:canonicalcuts}). The problem with this approach is that it is verbose and results in an unintuitive representation of the object. Shapiro and Vossler \cite{Shapiro1991cad} noticed the similarities between minimizing a CSG representation constructed with Algorithm \ref{brep2csg1}, and the problem of minimizing switching functions (also known as Boolean function minimization). 
In Boolean logic, an {\em implicant} is a product term (or conjunction of literals) whose truthfulness implies the truthfulness of a Boolean function. In our geometrical setting, an implicant is an intersection of halfspaces that is entirely contained within the solid to be represented. 
An implicant is {\em prime} if we can not delete any halfspace literal from the implicant. I.e. if we delete any halfspace litteral from the implicant, the result is not contained within the solid to be represented anymore (and thus is not an implicant anymore). 
Minimizing a CSG expression is equivalent to finding a prime implicant cover for the solid. This is a difficult problem (NP-complete). Shapiro and Vossler proposed in \cite{Shapiro1991cad} to use approximation algorithms to find a minimal cover, i.e. algorithms that compute an approximate minimal cover of the prime implicants. 

We can do better by noticing that in the best case each halfspace or its complement appears once in the CSG representation. A \emph{dominating halfspace} is a halfspace that is entirely contained within the solid, or in the complement of the solid. Factoring dominating halfspaces lead to a simplification of the CSG expression. This corresponds to the DTE representation (\ref{eq:dec}) introduced above, and it leads to Algorithm \ref{brep2csg2}. 
\begin{algorithm}
\caption{B-rep to CSG conversion via dominant primitives factoring}\label{brep2csg2}
\begin{algorithmic}
\STATE Factor dominating halfspaces of the solid. 
\IF{the resultant is empty (or the full set)}
\STATE Return // We have a CSG expression
\ELSE 
\STATE Iteratively try to perform a decomposition of the resultant. 
\STATE If the resultant is not decomposable, compute an approximate minimal cover of the prime implicants. 
\ENDIF 
\end{algorithmic}
\end{algorithm}
Buchele and Crawford proposed an improved version of this algorithm in \cite{Buchele2004}. Instead of computing an approximate minimal cover of the prime implicants when the resultant is not decomposable, they proposed instead to split the resultant by one of the halfspaces, and to apply the decomposition on each resulting subsets. 

\subsection{Range image to CSG}
Surprisingly, earlier works tried to recover a CSG expression directly from range images of an object \cite{Lin1988,Chen1992}. Range images were first segmented and implicit quadric surfaces were fitted to each subset. Similarly to many other systems, objects were assumed to be made of cubes, spheres, cylinders, cones and ellipsoids. The fitted primitives were determined to be positive or negative, and then combined by union or subtraction with their adjacent neighbors using a precedence graph. They made the strong (and limiting) assumption that each primitive could be subtracted at most once.

\section{CSG extraction from unstructured data}\label{sec:methods}
\begin{figure}[!h!t!bp]
\centering
\includegraphics[width=0.95\linewidth]{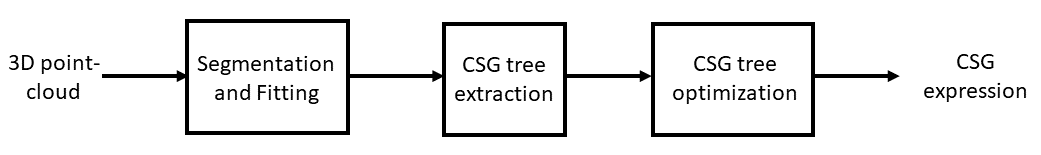}
\caption{Typical pipeline for extracting a CSG expression from an input 3D point-cloud.}\label{fig:pipeline}
\end{figure}

\begin{figure}[!h!t!bp]
\centering
\includegraphics[width=0.25\linewidth]{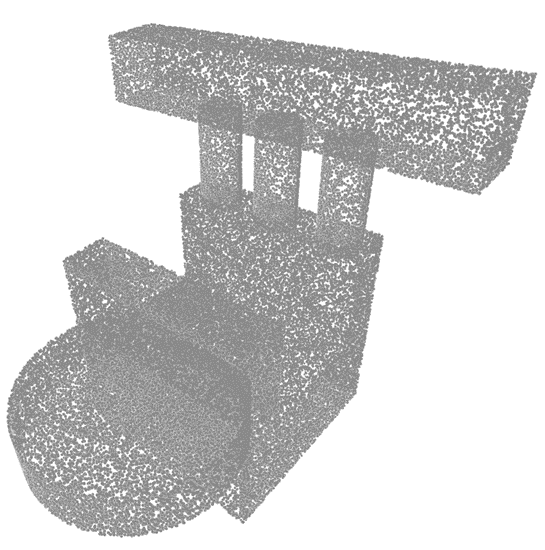}
\includegraphics[width=0.3\linewidth]{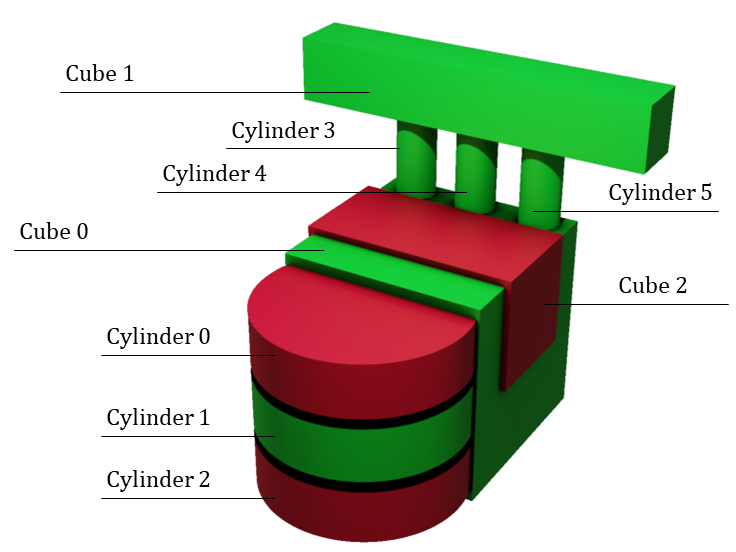}
\includegraphics[width=0.3\linewidth]{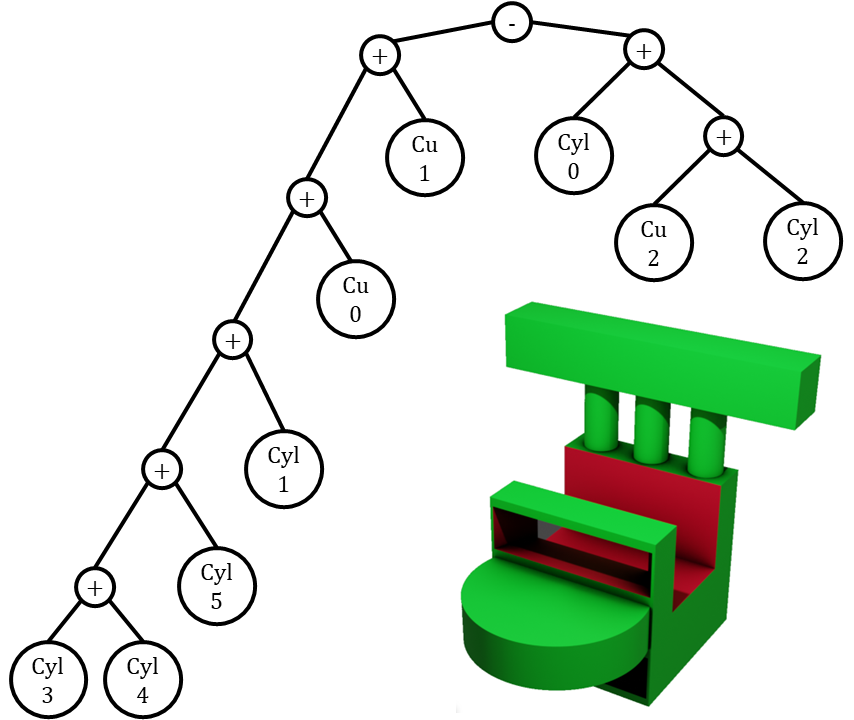}
\caption{An instance of the pipeline from Fig.\,\ref{fig:pipeline}. Left: The input point-cloud; Middle: The segmented primitives; Right: The optimized CSG tree / CSG expression and the corresponding solid.}\label{fig:pipeline_visu}
\end{figure}

We turn our attention now to the problem of recovering a CSG model from an unstructured input data set (3D point-cloud, triangle mesh or triangle soup). The usual pipeline for extracting a CSG expression from a 3D point-cloud is illustrated in the diagram shown in Fig.\,\ref{fig:pipeline} and an example is shown in Fig.\,\ref{fig:pipeline_visu}. This pipeline was introduced in \cite{Fayolle2008automation}. The input 3D point-cloud is first segmented into subsets and primitives are fitted to each of these subsets. The segmentation and primitive fitting step is difficult by itself and is the topic of several papers. See, for example, the recent survey \cite{Kaiser2019}. We will also discuss some of the available fitting approaches in Section~\ref{sec:fitting}. The next step deals with the extraction of a CSG expression, involving the primitives from the previous step, that describes the object corresponding to the input 3D point-cloud. Finally, an optional step tries to further optimize the CSG expression. Interestingly, this pipeline is not always used, and some approaches try to discover the primitives at the same time as they build the CSG tree, merging together the different steps of the pipeline. Examples of such approaches are described in \cite{SFVPRT05,Sharma2018csgnet,Ren2021ICCV}, among others. 

\subsection{Primitive fitting and shape decomposition}\label{sec:fitting}
\begin{figure}[!h!t!bp]
\centering
\includegraphics[width=0.95\linewidth]{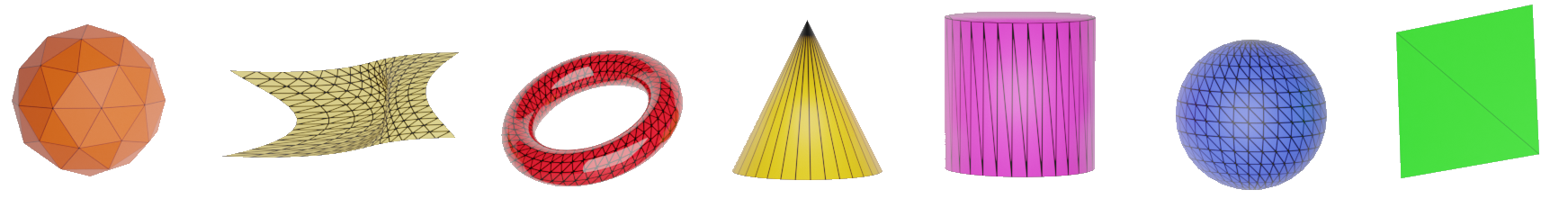}
\caption{A list of the geometric primitives typically used in CSG recovery.}\label{fig:primitives}
\end{figure}

The first step in the reconstruction pipeline consists in the segmentation of the input 3D point-cloud and the fitting of geometric primitives to the different subsets. 
Figure \ref{fig:primitives} shows a list of the primitives typically considered when recovering CSG expressions, and used as a candidate list for the fitting and segmentation step. 
The collection of the fitted geometric primitives can then be considered for the CSG tree extraction step. 
Some approaches stop after this first step, and simply try to represent the solid as the union of the geometric primitives (\ref{eq:canonicalcutssimp}). This is possible when the fitted primitives are sufficiently high-level such as in \cite{genova2019learning,paschalidou2019superquadrics}. 

\subsubsection{CAD-based segmentation and fitting}
Techniques for segmentation and fitting in reverse engineering and CAD are usually classified as bottom-up or top-bottom approaches \cite{VMC97}. 
The former use a region growing technique approach, starting from seed points. One example of such a method is described in \cite{BMV01}, where a system for reconstructing a B-rep from an unorganized point-cloud is described. Problems of bottom-up approaches include the difficulty to select seed points and the difficulty to decide whether to add points in a given region due to the presence of noise. 
Top-bottom approaches, while very popular in image segmentation, are less commonly used for surface segmentation task. The main difficulty is to find good criteria for deciding where and how to subdivide the input point-cloud. 

A different technique consists in computing the region boundaries from a network of feature curves. Usual techniques for surface fitting, based on least-square fitting, can be used on each region. This is the approach proposed in \cite{VFT07} where Morse theory is used on a triangulation of the input point-cloud to compute these feature curves. Algorithms for least-square fitting several common primitives (spheres, cylinders, cones and torii) were proposed in \cite{MLM01}. 

Another family of techniques is based on fitting more complicated parametric primitives to a point-cloud. This involves optimizing non-linear functions by non-linear least squares methods (such as, for example, Levenberg-Marquardt). This approach was applied, for example, to the domain of industrial engineering \cite{Rabbani2006}. 

\subsubsection{RANSAC based approaches}
RANSAC is an approach often used for the segmentation and fitting step \cite{fischler1981}. Its main advantage is its ability to deal with noise in the data. In geometric computing, the efficient RANSAC variant described in \cite{Schnabel2007} is the most popular. The main idea behind RANSAC is simple: Pick a few points (sufficient to define a geometric primitive), directly compute the parameters of the corresponding primitive, check how many points from the input point-cloud are sufficiently close to this primitive, and eventually accept the primitive based on statistical considerations. The efficient RANSAC approach of Schnabel et al. \cite{Schnabel2007} is using additional heuristics to accelerate this process when applied to a 3D point-cloud and 3D geometric primitives. 
The addition of graph-based constraints in \cite{li2011} to the efficient RANSAC approach is trying to improve the robustness of fitting of RANSAC by considering additional constraints such as, for example, the parallelism of planes, or the orthogonality of planes, among others. 

\begin{figure}[!h!t!bp]
\centering
\includegraphics[width=0.3\linewidth]{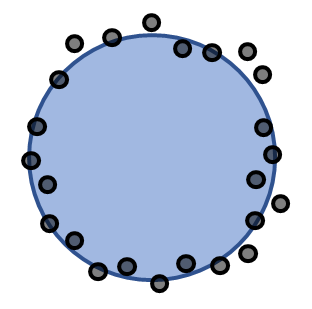}
\includegraphics[width=0.3\linewidth]{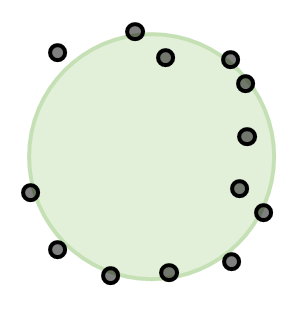}
\includegraphics[width=0.35\linewidth]{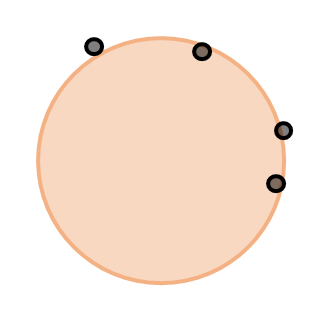}
\caption{One run of RANSAC. Some of the points are not sufficiently close to the fitted circle (left-most) because of the noise. They are later fitted by different, but close, circles (middle and right).}\label{fig:ransaconerun}
\end{figure}

\begin{figure}[!h!t!bp]
\centering
\includegraphics[width=0.3\linewidth]{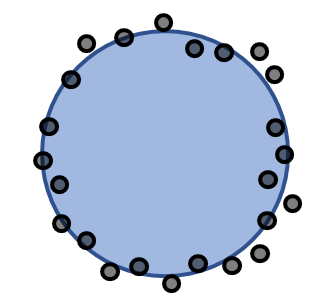}
\includegraphics[width=0.3\linewidth]{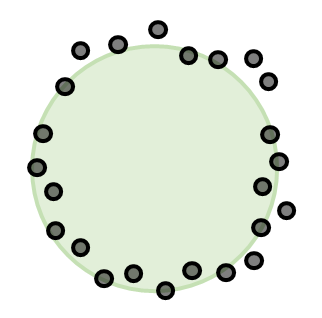}
\includegraphics[width=0.3\linewidth]{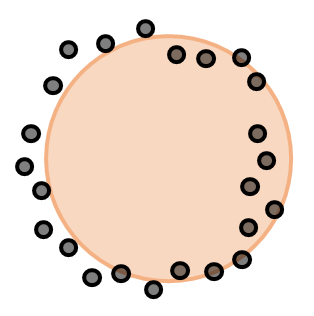}
\caption{A noisy point-cloud and several close circles segmented and fitted by RANSAC.}\label{fig:ransac}
\end{figure}

Despite its popularity, RANSAC-based methods suffer from a few problems: Their dependence on a few and sometimes unintuitive parameters (different parameters result in different primitives/parameters) or the fact that the method is stochastic (and thus, different runs will produce different primitives/parameters). This is illustrated in 2D in Fig.\,\ref{fig:ransaconerun} and  Fig.\,\ref{fig:ransac}.
Figure \ref{fig:ransac} corresponds to several iterations of one of RANSAC. Given the first fitted circle, some of the points are not sufficiently close and left. They are fitted to different, but close, circles in subsequent steps (middle and right images). Note that the use of slightly different parameters (such as the minimum support of points, or the distance threshold for matching points to a primitive) may produce different results. 
In Fig.\,\ref{fig:ransac} a noisy input point-cloud is segmented and fitted by RANSAC, each image corresponds to a different run of RANSAC with slightly different parameters. Here again, the produced output is a collection of close circles. In practice, one can run RANSAC multiple times with slightly different parameter values (sampled from a given distribution) and merge close primitives. Two primitives are considered to be close, if their defining parameters are within a given threshold. Another approach for improving the output of RANSAC is to try to discover and to enforce constraints between the primitives (parallelism, perpendicularity, ...). This approach was considered in \cite{li2011}.

\subsubsection{Deep learning based approaches}\label{sec:deep_fitting}
\paragraph{Deep learning based approaches for segmentation and primitives fitting}
The dependence of RANSAC on multiple parameters was one of the main motivations behind the supervised fitting approach described in \cite{Li2019}. The approach is based on combining traditional fitting methods with a deep neural network based on Point-Net++ \cite{pointnet2} for fitting primitives to point-clouds. Fully connected layers are added to Point-Net++ to predict attributes for each input point (the point-to-primitive membership, the surface normal at the point and the type of primitive that the point is assigned to). The parameters for each primitive are estimated by differentiable fitting methods based on these point properties. A similar approach for point-cloud segmentation and primitive fitting was described in \cite{yan_hpnet}. The proposed approach, called HPNet, consists in learning several descriptors (semantic descriptors, spectral descriptors, sharp edge descriptors) and learning the combination weights of the descriptors. 
A strong limitation of such approaches is the size of the input point-cloud, which is limited to a few thousand points. This constraint was later addressed in \cite{le2021cpfn} by the inclusion of a selection network. 

An extension from the usual quadrics to parametric surfaces (B-Splines patches) was considered in \cite{sharma2020parsenet}. A neural network is used to classify points from the input point-cloud into different classes corresponding to the type of primitives (plane, cylinder, B-Spline, ...). Quadrics are estimated by least-square fitting, while the B-Splines are fitted by another neural network. 
The segmentation and fitting of generalized cylinders (a sweep of a 2D domain along a given curve) was proposed in \cite{uy2021point2cyl}. The approach is based on combining neural networks for the segmentation, estimation of some quantities (such as the surface normals) and base curve fitting, and solvers for computing the extrusion cylinder parameters. All these approaches rely on supervision (or at least partially). 

Most of the previously described methods rely on supervised learning, i.e., the availability of a ground truth for training the models. In practice, these methods are trained on certain distributions of objects, collected into large data sets (see Section~\ref{sec:dataset}), and tend to perform poorly for point-clouds coming from different distributions. In order to address this shortcoming, AutoGPart \cite{liu_autogpart} was introduced. It works by building a supervision space using encoded geometric priors, and by searching the optimal supervision from that space. 

\paragraph{Deep learning based approaches for shape decomposition and shape approximation}
Neural network based approaches have also been used for the problem of shape decomposition. Most of these approaches are unsupervised. Often, they target an approximation of the shape only. These methods are related to the union expression form (\ref{eq:canonicalcutssimp}) and do not need to pass through the CSG generation step (The CSG expression is the union of all the primitives). In general, these approaches deliver an approximation only, they often do not retrieve sharp features, and do no not necessarily result in a usable (editable) expression. 

An approach for learning to approximate (or to abstract) shapes from a collection of simple volumetric primitives (box primitives) was introduced in \cite{tulsiani2018learning}. The approach is limited to a (fixed) small number of box primitives (3 to 6). The related method, introduced in \cite{Yang2021}, proposed to learn a cuboid shape abstraction by unsupervised learning via joint segmentation from point-clouds. Another possible generalization is to consider the decomposition into a collection of convex polytopes, as proposed in \cite{deng2020cvxnet}. 
Replacing box/cuboid/convex polytope by more general superquadric primitives was proposed in \cite{paschalidou2019superquadrics}.
Learning a general shape template from data by collections of ellipsoids/Gaussians was proposed in \cite{genova2019learning}. 
Using simple primitives lead to a crude approximation of the shape, in \cite{paschalidou2021neural} the authors addressed this problem by considering a different representation for the primitives (invertible neural network). This is done at the expense of the size of the representation. Another possible solution was proposed in \cite{genova2020local} with the use of local implicits. 

\subsubsection{Separating primitives}
Similarly to the case of the B-rep to CSG conversion approaches, the set of primitives fitted to a surface may not be sufficient to describe the corresponding solid and form a well-defined CSG representation. See the discussion in Section~\ref{sec:describability} and Fig.\,\ref{fig:halfspaces}, right image. The solution consists in adding extra primitives to the set of primitives fitted to the point-cloud (or triangle mesh). It seems natural to consider the planes corresponding to the bounding box of the points associated to a given non-planar primitive \cite{Fayolle2016evolutionary}. The bounding box can be aligned with the axes of the coordinate system or be aligned with the main directions of the point-cloud (obtained by principal component analysis). 

The addition of such separating planes may not be sufficient and one can consider adding other types of separating primitives, such as adding torii for each cylinder, or more general canal surfaces \cite{Fayolle2016evolutionary}.

\subsection{CSG generation}\label{sec:generation}
We turn our attention to the problem of CSG generation. Our input is an unstructured point-set (typically a point-cloud, or eventually a triangle mesh) and a list of primitives made of the primitives fitted to the point-cloud and the additional separating primitives. The problem is to combine these primitives with CSG operations such that the corresponding solid describes the domain associated to the input point-cloud. This is a difficult combinatorial search.

\subsubsection{Boolean logic based approaches}
The first possible approach to this problem is to adapt the methods for B-rep to CSG conversion described in Section~\ref{sec:brep2csg}. Namely, we would like to use Algorithm~\ref{brep2csg1} or Algorithm~\ref{brep2csg2} or the variant based on halfspace splitting \cite{Buchele2004}. 

For this purpose, we need to build a Point-Membership-Classification (PMC) function for the point-cloud. It will be used for classifying the fundamental products (see Algorithm~\ref{brep2csg1}). 
The typical approach for obtaining a PMC function consists in first performing surface reconstruction from the point-cloud (using, for example, \cite{Kazhdan2006}). If an implicit surface is fitted to the point-cloud, then it can be used for the PMC queries, otherwise traditional methods for point classification w.r.t. a triangle mesh can be used \cite{Barill2018}. 
An alternative approach consists in voxelizing the point-cloud. Here, we are interested in the voxelization of the volume bounded by the surface from which the point-cloud was sampled (solid voxelization). Once again, it requires to first perform surface reconstruction of the input point-cloud, and then to voxelize the domain bounded by the triangle mesh. 
These additional numerical computations can potentially be fragile, or may introduce additional approximation in the data. Nonetheless, they are necessary for adapting the algorithms for converting B-rep models to CSG models, as well as for the implementation of several of the algorithms that we will look at below. 

The approach used by Xiao and Furukawa for the automatic construction of museums \cite{xiao2012,xiao2014} can be seen as related to the Boolean logic based approaches used in B-rep to CSG conversion. Their approach consists in precomputing all possible cuboids (the primitives) and greedily add them (which corresponds to a union) or remove them (which corresponds to a difference). To accelerate the computations, the approach is decomposed in two steps: In the first step, the approach is carried in two-dimensions (the primitives are rectangles). The resulting 2D CSG model is then used to generate the candidate 3D cuboid primitives, limiting the combinatorial explosion. The candidate cuboids are then greedily added or removed. An objective function, measuring volumetric fitting and surface fitting, is used to guide the search. 

A similar approach based on Boolean logic and heuristics is proposed for reconstructing CSG models from 3D point-cloud \cite{Wu2018}. This approach is not limited to planes but uses all of the typical quadric primitives (planes, spheres, cylinders, torii, cones). It uses the typical pipeline consisting of point-cloud segmentation and fitting (using RANSAC), followed by CSG recovery. The CSG recovery stage shares similarity with the B-rep to CSG approach of Shapiro and Vossler \cite{Shapiro1991jmd}, but with added heuristics to perform the combinatorial search. As an additional optimization, they combine planes into cuboids, and clip the other quadric primitives by taking the intersection with their oriented bounding box in the primitive fitting step. This optimization allows to simplify the combinatorial CSG search. 

The method described in 
\cite{friedrich2022full} uses as well a full pipeline for recovering a CSG model from a point-cloud. The segmentation and solid primitive fitting steps are a variant of the approach described in \cite{friedrich2020grapp}. The CSG recovery step is a variant of Algorithm~\ref{brep2csg2}, where the dominating primitives are first factored from the solid, and the resultant solid is computed by a Genetic Algorithm (GA). Two different strategies are used for the GA: A selection based GA, and a node based GA. In the selection based GA, an individual is represented by a bit-string, with one bit associated to each remaining primitive. Each added primitive is either added or removed, depending on its classification. On the other hand, the node based GA is evolving a CSG sub-tree for the resultant solid, using techniques similar to the ones presented below in Section~\ref{sec:evolutionary}.

\subsubsection{Evolutionary based approaches}
\label{sec:evolutionary}
\begin{figure}[!h!t!b!p]
\centering
\includegraphics[width=0.95\textwidth]{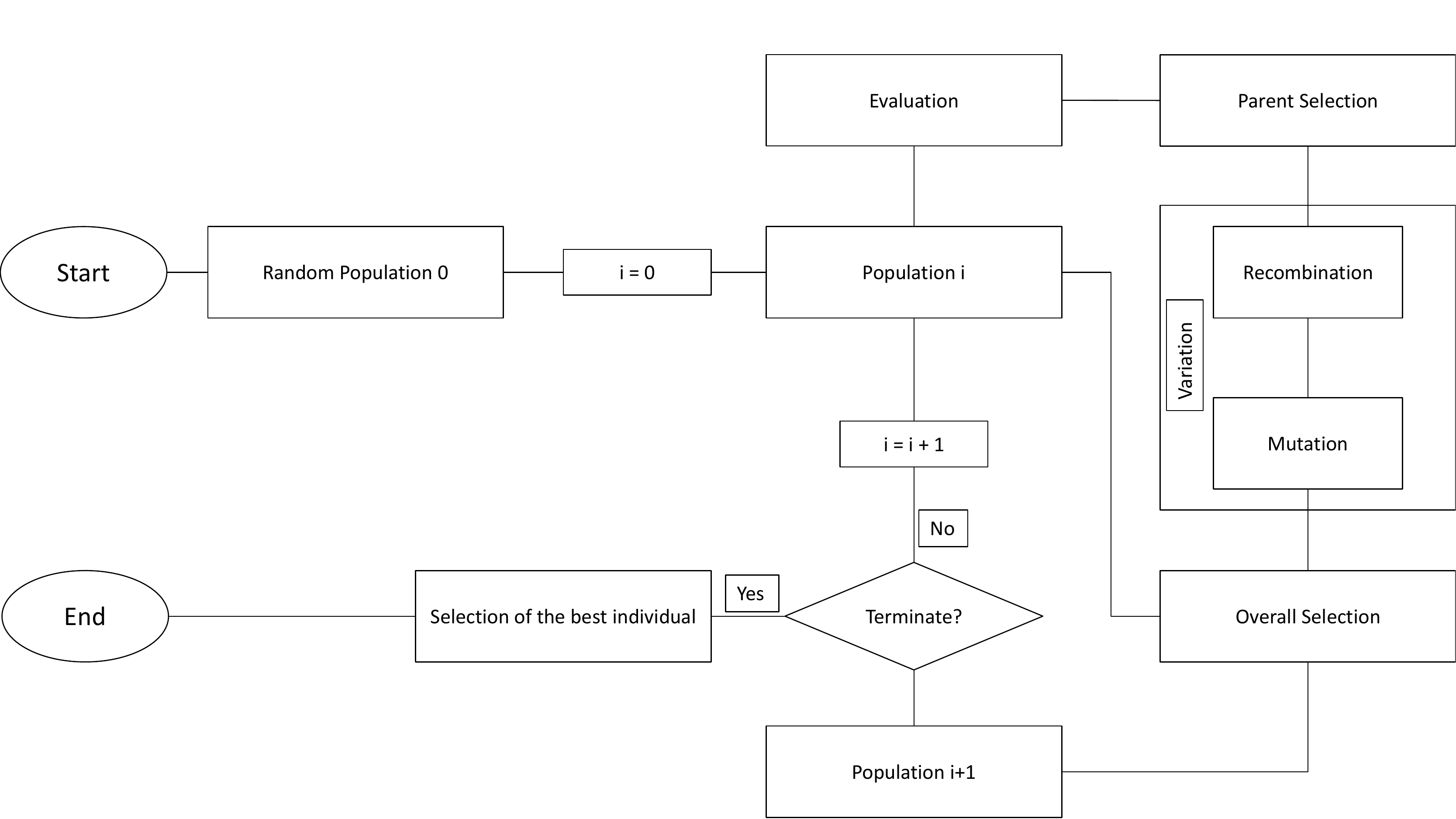}
\caption{Diagram illustrating a typical evolutionary algorithm.}\label{fig:ea_diagram}
\end{figure}

Evolutionary computation is an ensemble of optimization techniques inspired by biological evolution. These techniques are useful for approximating the global optimum to difficult optimization problems. A typical evolutionary algorithm is illustrated with the diagram shown in Fig.\,\ref{fig:ea_diagram}. The problem under consideration consists in recovering a CSG expression (or, equivalently, a CSG tree) corresponding to a given sampling of an object. It makes sense to consider Genetic Programming (GP) for this purpose. Genetic Programming is an evolutionary technique invented by John Koza for evolving computer programs (trees) satisfying a criterion encoded by an objective function \cite{Koza92}. 

One of the first works using GP to evolve a CSG tree from a point-cloud was the work of Silva et al. \cite{SFVPRT05}. Unlike the pipeline shown in Fig.\,\ref{fig:pipeline}, they used a GP to automatically evolve both the construction (CSG) tree and the parameters of the primitives at the same time. This particular approach made it impossible to use a standard GP, a strongly-typed GP was used instead \cite{montana1995}. Two types are considered: The first corresponds to the primitives (cylinder, cuboid, ...) and CSG operations; The second type corresponds to the parameters of the primitives. To limit the complexity of the generated CSG expressions, which is a known problem of GP-based approaches, parsimony pressure measures are used \cite{Koza92,luke02,silva09}. Unfortunately, the experimental results obtained by this approach were rather limited. Evolving at the same time the parameters of the primitives and the CSG expression is too much work, and it seems more reasonable to tackle each problem separately. 

The first appearance of the typical pipeline: point-cloud segmentation, fitting, and CSG recovery, shown in Fig.\,\ref{fig:pipeline} was in \cite{Fayolle2008automation}. The segmentation of the input point-cloud and fitting of primitives to the different clusters was done by a Genetic Algorithm (GA). To prevent the CSG expressions to grow unbounded, they decided to use a simple GA to select which primitive to use and the CSG operation to apply on the primitive, i.e., whether the primitive should be added or subtracted to the representation, or whether the intersection with the solid built so-far should be taken. The main problem of this GA-based recovery approach is that it will only work with solids that are fully decomposable. This GA approach is closer in spirit to some of the Boolean logic approaches described above. 

\begin{figure}[!h!t!bp]
\centering
\includegraphics[width=0.3\linewidth]{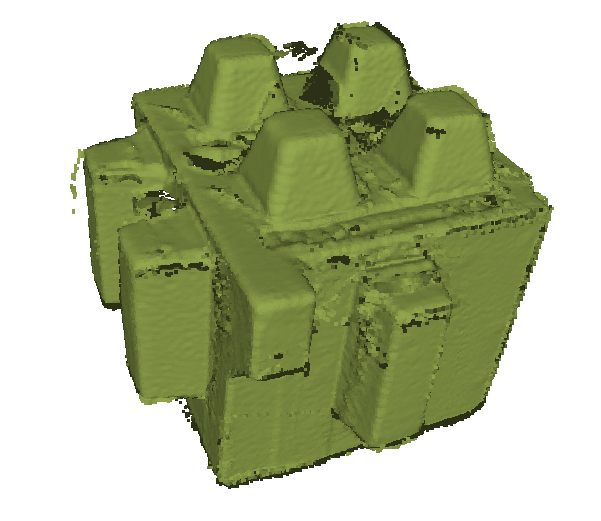}
\includegraphics[width=0.3\linewidth]{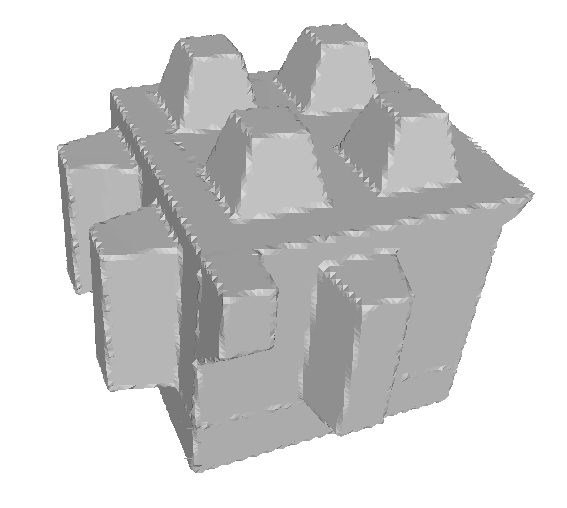}
\includegraphics[width=0.27\linewidth]{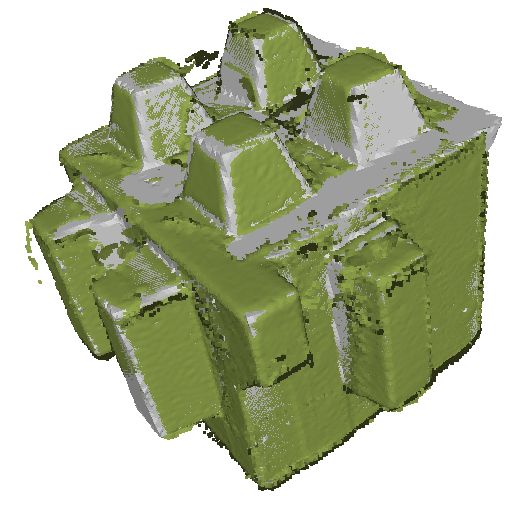}
\caption{Example of a complex and noisy point-cloud (left image) and the corresponding CSG model recovered by the method described in \cite{Fayolle2016evolutionary} (middle image). Right: For visualization purpose, both the point-cloud and the cSG model are shown together. The CSG model was converted to an implicit model (See Section \ref{sec:csgsdf}) and meshed by the Marching Cubes algorithm.}\label{fig:gp4csg}
\end{figure}

Fayolle and Pasko kept a similar pipeline but proposed to replace the simpler GA by a full fledged GP \cite{Fayolle2016evolutionary}. This addresses the main limitation of the approach \cite{Fayolle2008automation}. To prevent the CSG trees evolved by GP to grow unbounded, they rely on a simple heuristic: A term penalizing deep trees is included in the objective function optimized by GP. Another major difference with \cite{Fayolle2008automation} is the introduction of a step for computing separating primitives. The approach was shown to recover CSG models from noisy and complex point-clouds. An example is shown in Fig.\,\ref{fig:gp4csg}. This approach has two problems: First, it is relatively inefficient. Genetic Programming works by maintaining a population of creatures, each of these creatures represents an expression whose fitness should evaluated at each iteration. In the case of \cite{Fayolle2016evolutionary}, the creatures are CSG expressions (potentially large), and evaluating their fitness means evaluating these CSG expressions at each point of the input point-cloud. While caching mechanisms can be implemented to avoid un-necessary re-computations, the method still remains inefficient. The second problem of the approach is that it does not make any use of spatial information about the primitives, so neighbor primitives may eventually be located far away in the CSG tree, making it difficult to reason about the CSG tree or edit it. 

\begin{figure}[!h!t!bp]
\centering
\includegraphics[width=0.9\linewidth]{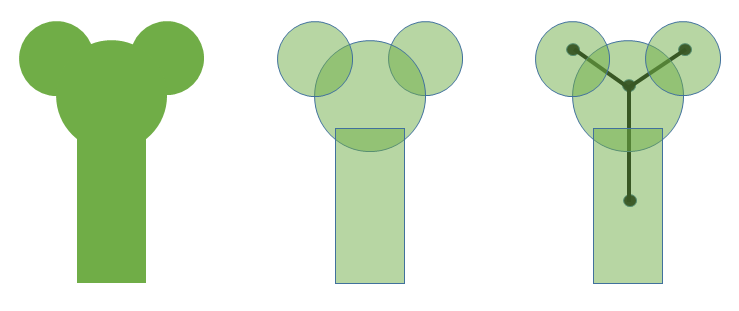}
\caption{Graph partitioning for accelerating evolutionary approaches. Left: The solid to be recovered. Middle: The different primitives defining the solid (fitted by RANSAC). Right: The primitive intersection graph with one vertex per primitive and one edge between two intersecting primitives.}\label{fig:graph_partitioning}
\end{figure}

Both problems were addressed in subsequent works by Friedrich et al. by using techniques from graph partitioning \cite{Friedrich2018accelerating,Friedrich2019gecco}. Given a collection of fitted primitives, see Fig.\,\ref{fig:graph_partitioning}, the primitive intersection graph is built by adding one vertex per fitted primitive and one edge between two intersecting primitives. One possible approach to partition the primitive intersection graph consists in computing all the maximum cliques (the maximum complete subgraphs), a GP-based CSG tree extraction is then independently run on each partition (each maximum clique), finally the CSG expressions corresponding to each partition are merged to form the final CSG tree \cite{Friedrich2018accelerating}. This approaches has two main problems: Computing the maximum cliques of a graph is a difficult task; And the processing for merging CSG trees is complicated. 
To resolve these problems, a different partitioning scheme was used in \cite{Friedrich2019gecco}: It is still based on partitioning the primitive intersection graph, but instead of computing the maximum cliques, it relies on pruning certain types of primitives and disconnecting the prime implicants primitives (by removing the corresponding edges in the graph), and then computing the connected components of the graph. A GP is run independently (and potentially in parallel) on each connected component. The final CSG expression is simply the union of the CSG expressions computed for each partition. 
In addition to the processing speed, one of the main advantages of these partition-based approaches is that the primitives in each partition are spatially close, which result in CSG trees that are more intuitively editable (the sub-trees are involving spatially close primitives).

\subsubsection{Program synthesis based approaches}
\label{sec:prog_synth}
Genetic Programming can be viewed as one particular type (stochastic search) of a more general approach called program synthesis. For an introduction to the topic of program synthesis, the reader can refer to the following recent survey \cite{gulwani2017program}. Program synthesis techniques are methods for generating computer programs (usually expressed in a given domain specific language) that meet some given specifications. In our case, the goal is to generate a CSG program, consisting in simple primitives and Boolean operations, from a given specification, which is an unstructured representation of an object (such as a point-cloud, a bitmap image or a voxel representation). In the recent years, there has been an increased interest from the programming language community for this problem of CSG recovery (sometimes under the term of CAD de-compilation). 

Du et al. proposed to combine tools from geometry processing and program synthesis for tackling the problem of de-compiling a 3D triangle mesh to a CSG expression \cite{Du2018}. Rather than relying solely on techniques from program synthesis, they relied on a pipeline similar to \cite{Fayolle2016evolutionary} but with the stochastic search (GP) replaced by a different technique for program synthesis called "sketching" \cite{SolarLezama2008}, and several other improvements for the different steps of the pipeline. In particular, primitive detection is done with the efficient RANSAC method of Schnabel et al. \cite{Schnabel2007}, combined with graph-cut \cite{Boykov2004} and other heuristics to improve the primitive fitting part (such as, for example, the combination of pairwise orthogonal/parallel planes into cuboids). Generating a CSG expression for a large number of primitives by sketching is un-tractable, so they sub-divide the input shape into smaller (involving fewer primitives) and more tractable shapes. 

Similar approaches, based on techniques and tools from program synthesis are used in \cite{Nandi2017,Nandi2018,Nandi2020,Nandi2021} to decompile low-level triangle meshes to CSG expressions. The main approach is search-based similarly to \cite{Du2018}. However, unlike the approach of Du et al., it does not rely on geometric algorithms for identifying the primitives and their parameters, but rely solely on program synthesis techniques. In order to guide the program synthesis, evaluation context is used, which leads to a more efficient navigation of the search space of possible CSG programs. CAD models often exhibit repetitive features or patterns. In order to properly recover these repetitive patterns, Nandi et al. use a post-processing technique to capture such repetitions and extract loops \cite{Nandi2020}. However, this approach is potentially expensive, because it requires to extract first a program without any loop. 

Feser et al. proposed a different approach for the CSG recovery problem based on the concept of metric program synthesis \cite{Feser2022}. Their method is based on the observation that several CSG models will produce similar, but not identical, solids. From this observation, they design an algorithm for generating CSG programs that approximately match the input, and then improve on these CSG programs by a local refinement based on tabu search \cite{Glover1998}. While their approach is a more sophisticated synthesis algorithm that can handle a \emph{repeat} operator for dealing with repetitive patterns, their experiments are limited to simple 2D bitmap images, and only involve disks and rectangles as primitives, and union and difference as operations.

\subsubsection{Deep learning based approaches}
Deep learning based techniques have become increasingly popular tools for solving problems in different domains of application. Originally, they were targeting problems in computer vision and natural language processing, but their domains of applications have expanded since then. 
We have already seen in Section~\ref{sec:deep_fitting} several ways in which deep neural networks could be used for the problems of segmentation of a 3D point-cloud and primitives fitting. 
We are now interested in CSG expression generation and, thus, problems related to combinatorial optimization and language generation. Traditionally, combinatorial optimization is not a domain where deep neural networks have shun, however, a recent work \cite{Schuetz_2022} has shown how to use graph neural networks \cite{Bronstein_2017,Bronstein_2021} to solve combinatorial optimization problems, and could lead to interesting directions. 

On the other hand, natural language processing is a domain where deep learning solutions have had a lot of successes and it seems legitimate to look at first in this direction for generating CSG expressions. For example, one could use a recurrent neural network to generate a simple CSG expression. This approach was considered in CSG-Net \cite{Sharma2018csgnet,Sharma2020}. The authors used a deep neural network for converting an image in 2D (collection of pixels) or voxels in 3D into a CSG expression corresponding to the input shape. Their approach is based on a convolutional layers for encoding the input image into a low-dimensional latent space, and a recurrent neural network to decode the latent vector into a CSG expression. The training is done by supervised learning when the ground truth is available, or by policy gradient techniques otherwise. CSG-Net does not rely on separate processes for primitive fitting and the CSG expression recovery, and as such, seems limited to very simple shapes. Additionally, it supports only fully decomposable models and thus solves a much simpler sub-problem ($O(n^2)$ instead of $O(2^n)$, where $n$ is the number of primitives). 

Somewhere in-between program synthesis techniques and deep learning methods are neural guided synthesis techniques (i.e., the use of deep neural networks to guide the program synthesis). An early example of neural guided synthesis for the CSG recovery problem was described in \cite{Ellis2019write}, where the output of partially constructed models was exploited to guide the search. Neural guided synthesis was also used for a more general CAD reconstruction in \cite{Willis2021}. A common problem to both approaches is that they require a lot of work to collect data sets and train the algorithms on these data sets.  

\begin{figure}[!h!t!bp]
\centering
\includegraphics[width=0.9\linewidth]{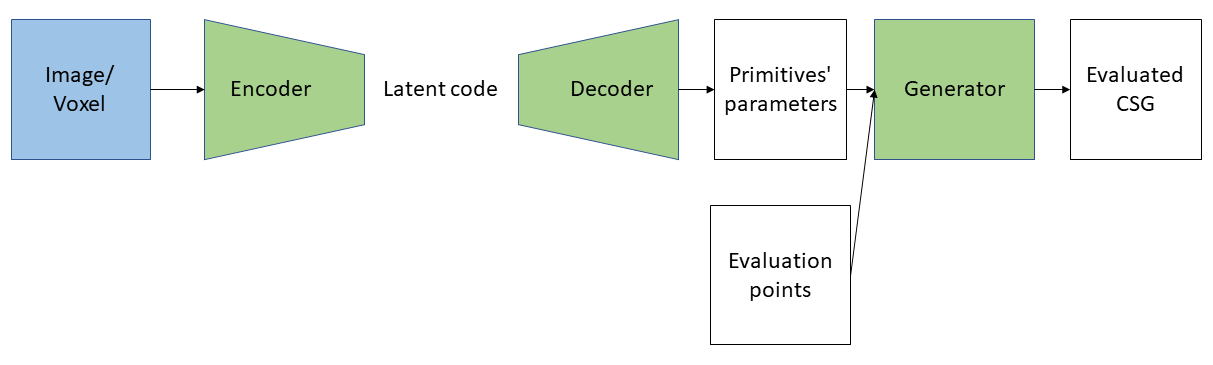}
\caption{General architecture of Cvx-Net \cite{deng2020cvxnet}, BSP-Net \cite{Chen2020bspnet}, CSG-Stump \cite{Ren2021ICCV} and CAPRI-Net \cite{Yu_2022_CVPR}. For Cvx-Net, the generator forms a fixed number of convex shapes and take their union. For BSP-Net, CSG-Stump and CAPRI-Net, the generator learns masks used to decide which primitives are combined by intersection and union.}\label{fig:deep_learning}
\end{figure}

Another possible direction of approach is to adapt the approach proposed by Shapiro and Vossler \cite{Shapiro1991cad} and based on the decomposition in canonical cells (\ref{eq:canonicalcuts}). The idea is to form, or "learn", convex cells and to take their union. Convex cells are formed by taking the intersection of primitives whose parameters are obtained from an auto-encoder. This approach was followed by Cvx-Net \cite{deng2020cvxnet}, BSP-Net \cite{Chen2020bspnet}, CSG-Stump \cite{Ren2021ICCV} and CAPRI-Net \cite{Yu_2022_CVPR}. In each of these methods (with the exception of CSG-Stump), the input is a bitmap image (in 2D) or voxels (in 3D), which is transformed into a latent code by an encoder (CSG-Stump assumes a 3D point-cloud and uses DGCNN \cite{Wang_dgcnn} as an encoder). This latent code is then decoded into the parameters of primitives: While Cvx-Net and BSP-Net are limited to planes, CAPRI-Net considers quadrics and CSG-Stump uses spheres, boxes, cylinders and cones. 
While Cvx-Net takes the union of all the convexes, the other approaches use trainable matrices that act as masks to decide if the primitive/convex should be part of an intersection/union. From a practical point of view, CAPRI-Net can be seen as a generalization of BSP-Net where planes are replaced by general quadrics. CSG-Stump and CAPRI-Net share a lot of similarities, with some minor differences in the formulation. All these approaches are trained (unsupervised learning) on collections of shapes from ShapeNet \cite{chang2015shapenet}. 

It is interesting to note that the authors of CSG-Stump propose also a different approach \cite[Section~3.3]{Ren2021ICCV}, which is not based on deep-learning, and shares a lot of similarities with some of the techniques described previously. This approach is based on fitting geometric primitives by RANSAC \cite{Schnabel2007}, followed by minimizing a binary programming problem, whose solution provides the binary masks for the intersection, complement and union operations. The main purpose of introducing a deep neural network is for dealing with the cases where the number of primitives is very large, and can not be handled in practice by binary programming solvers. 

Finally, one additional approach for unsupervised learning of CSG expressions from raw point-clouds is UCSG-Net \cite{Kania2020ucsgnet}. It combines an auto-encoder for extracting the primitives parameters with Gated Recurrent Units \cite{Cho_2014} for generating a parse tree in an unsupervised manner. The considered primitives are limited to boxes and spheres, and the results in 3D are rather limited. Similarly to the previous methods, it is trained (in an unsupervised manner) on ShapeNet. 

One of the possible problems of deep learning based approaches is that they require training on large collection of data sets (even if the training is unsupervised), which is time consuming. It is also not clear what the result will be when these methods are used to generate CSG expressions for out-of-distribution input data sets.

\subsubsection{CSG optimization}
Once a CSG expression has been obtained for a given unstructured input data, a final step consists in optimizing the expression. The first thing to clarify is what we mean by optimizing a CSG expression. Following Occam's razor, it seems reasonable to consider the size of the CSG expression as one criterion, where the size of a CSG expression is defined as the number of primitives and operations in the expression. It should be minimized. For example, we want to remove redundant primitives and operations (i.e. expressions such as $A \cup A$ should be replaced with $A$).

Assuming that the size of a CSG expression is optimal, another criterion worth considering is the \emph{editability} of the CSG expression (or CSG tree). For example, editing a sub-tree should result only in local modifications of the shape. We approximate the editability of a CSG expression by considering the \emph{spatial proximity}. It is defined as the ratio of the number of CSG operations whose operands overlap to the number of CSG operations whose operands are disjoint. We wish to maximize the spatial proximity of a given expression. 

\paragraph{Size optimization}
Optimization of a CSG expression was initially considered by Shapiro and Vossler in \cite{Shapiro1991cad}. Starting from a DNF expression (\ref{eq:canonicalcuts}), one can use techniques from switching theory \cite{quine1952,mccluskey1956,orourke1982} to minimize the expression. In Boolean logic, an \emph{implicant} is a product term, i.e. a conjunction of literals, that implies the truthfulness of the Boolean function. In our case, an implicant is an intersection of primitives (halfspaces), and the Boolean function is the CSG expression corresponding to a solid. An implicant is \emph{prime} when it can't be factored further (i.e. we can't remove any of its literals) \cite{quine1952}. In simple cases, the disjunction of all the prime implicants of a Boolean function gives the shortest possible disjunctive normal form of a function. Often, it is possible to do better, however finding the best disjunctive normal form is difficult in general and is NP-complete as it corresponds to an instance of the set cover problem. Shapiro and Vosler proposed to use algorithms that compute an approximate minimal cover of the prime implicants \cite{Shapiro1991cad}. 

In practice, the early factorization of the dominant halfspaces helps in limiting the size of the produced CSG expression \cite{Buchele2004}. An halfspace is \emph{dominant} if it is entirely contained inside or outside the target solid \cite{Shapiro1991cad,Buchele2004} (see also the DTE representation (\ref{eq:dec} in Section \ref{sec:csg}). 

Depending on the technique used, the produced CSG expression may contain repeated primitives or sub-expressions that are redundant in terms of the information that they provide for the solid description. In the domain of genetic programming, these are called \emph{introns} \cite{Koza92}. Removal of redundant sub-expressions can be done with an approach inspired from \cite{Tilove1984}. It consists in iteratively applying a set of simplifying rules to the CSG expression until none applies \cite[Section 5.1]{Friedrich2020optim}. 

Alternatively, when the CSG expression is generated by an evolutionary approach (see Section\,\ref{sec:evolutionary}), it is possible to introduce a regularization term in the objective (fitness) function that penalizes large trees (or expressions). This is the approach followed in \cite{Fayolle2016evolutionary}, for example. Instead, it is also possible to consider a multi-objective optimization problem, where one of the objective function optimizes for the geometric fidelity of the CSG expression and another optimizes for the size of the expression. This is the approach followed in \cite{Friedrich2019gecco}. 

\begin{figure}[!h!t!bp]
\centering
\includegraphics[width=0.3\linewidth]{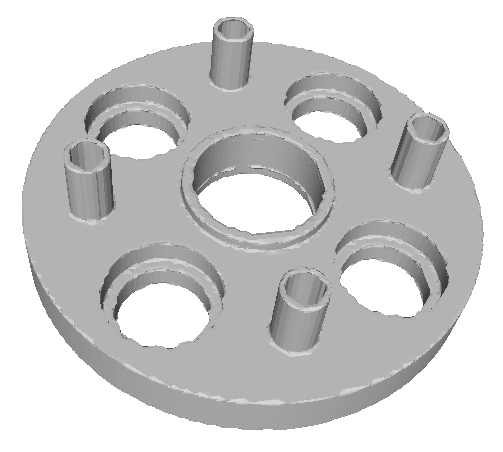}
\includegraphics[width=0.65\linewidth]{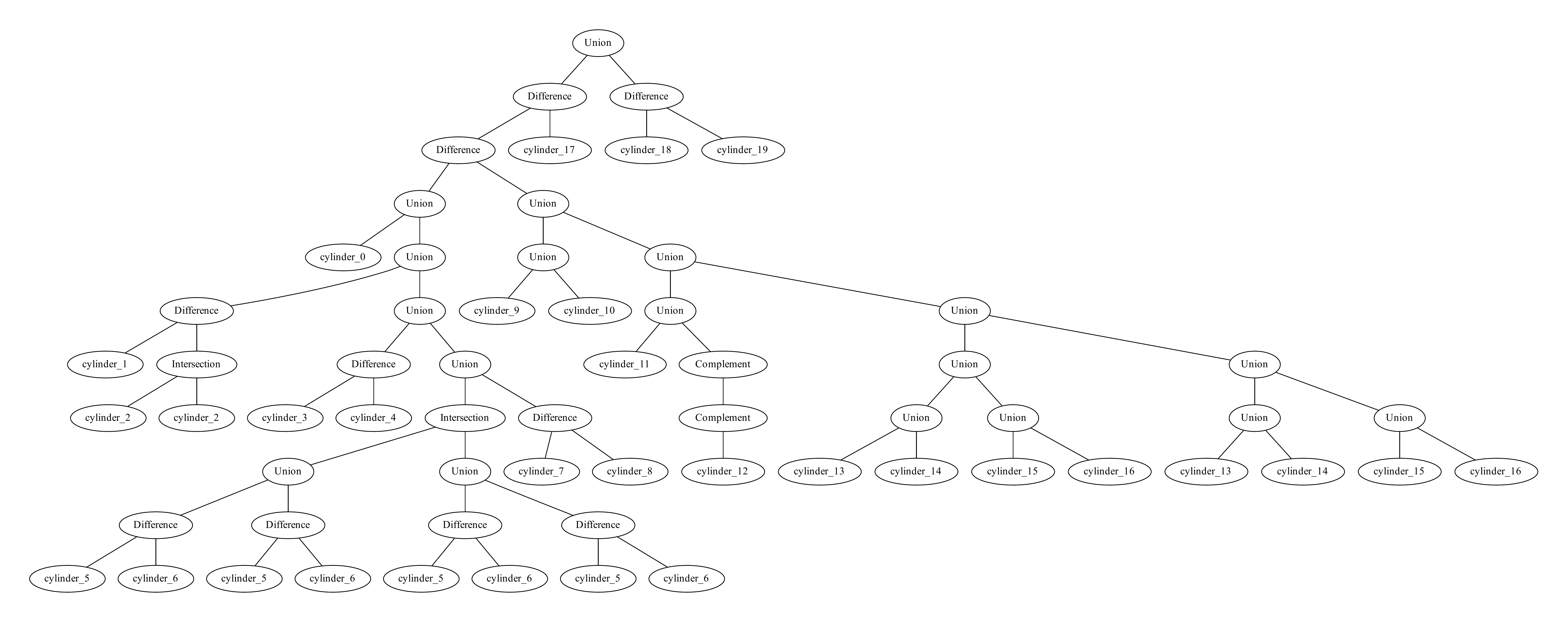}
\caption{Example of a CAD model and its CSG tree.}\label{fig:optim4csg}
\end{figure}

\begin{figure}[!h!t!bp]
\centering
\includegraphics[height=0.9\textheight]{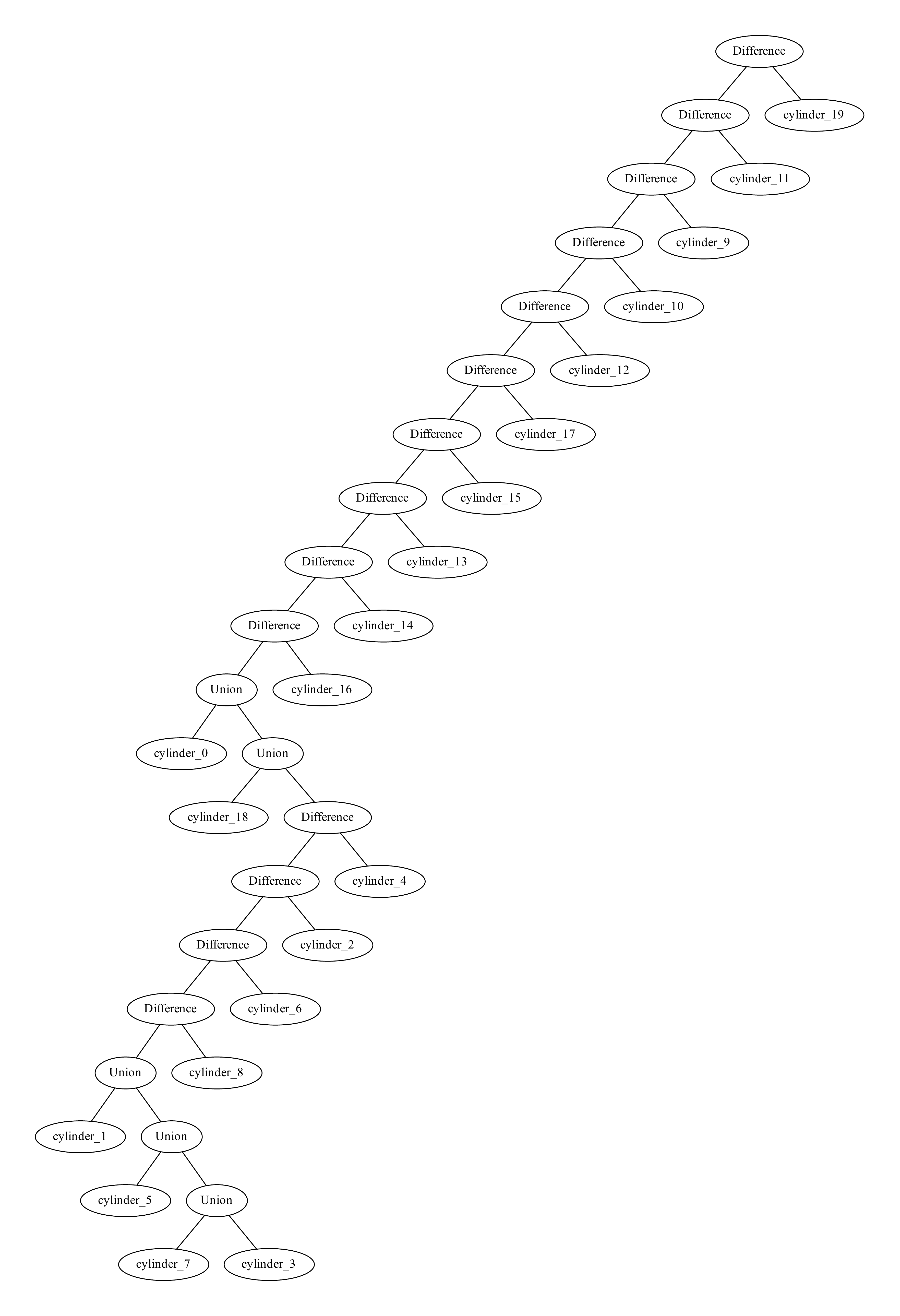}
\caption{CSG tree corresponding to Fig.\,\ref{fig:optim4csg} after optimizations.}\label{fig:optimized}
\end{figure}

Figures\,\ref{fig:optim4csg} and \ref{fig:optimized} illustrate an example of CSG tree optimization. A simple CAD model and its corresponding CSG tree is shown in Fig.\,\ref{fig:optim4csg}. The primitives and the CSG tree were recovered by some of the techniques mentioned earlier in Sections\, \ref{sec:fitting} and \ref{sec:generation}. The CSG tree consists in 31 primitives and 31 operations. The optimized CSG tree obtained by the techniques previously described is shown in Fig.\,\ref{fig:optimized}, and contains 20 primitives and 19 operations. 

\paragraph{Spatial proximity optimization}
Andrews proposes to simplify a CSG expression by removing spatially distant primitives from each fundamental product cell \cite{Andrews2013}, since these distant primitives are unrelated to the given fundamental product cell. This is related to the problem of \emph{editability} and the fact that editing a sub-tree of a given CSG model should result only in local modifications of the shape. When the CSG expression is generated by an evolutionary approach (Section\,\ref{sec:evolutionary}), it is possible to add a term to the objective function for enforcing that two operands of a Boolean operation are spatially connected. For example, one can consider the following term: 
$$
f(\Phi) = \frac{P(\Phi)}{\#|\Phi|},
$$
where $\#|\Phi|$ counts the number of nodes in the CSG tree $\Phi$, and $P$ is recursively defined by 
\begin{itemize}
\item $P(\Phi) = 1$ if $\Phi$ is a primitive, 
\item $P(\Phi) = P(\Phi_1) + P(\Phi_2) + \delta(\Phi_1, \Phi_2)$ when $\Phi$ is made of the sub-trees $\Phi_1$ and $\Phi_2$, and $\delta(\Phi_1,\Phi_2)=1$ if $\Phi_1$ and $\Phi_2$ intersect, $0$ otherwise.
\end{itemize}
This technique was used in \cite{Friedrich2020optim}.

\section{Discussion and conclusion}\label{sec:conclusion}
\subsection{Programs and higher-level representations}
We have seen earlier, in Section\,\ref{sec:prog_synth}, that a CSG expression can be seen as a program written in a simple language and that recovering a CSG model can be understood as an example of program synthesis. More generally, one can consider the problem of generating computer programs representing solids \cite{Ganin2021computeraided}, looking for example at higher-level representations \cite{Ellis2019write}. 

Program synthesis techniques for recovering a CSG expression are limited to combining simple geometric primitives (cubes, cylinders, \ldots) with Boolean operations (union, intersection, difference). However, such expressions are very primitives, and lack higher-level constructs, such as functions (grouping common functionalities) or loops, it makes it difficult to edit these CSG models. Using a technique called equality saturation \cite{Tate2009}, the work \cite{Nandi2020} proposes to augment the synthesis technique with map and fold operators. They argue that the generated CSG models (augmented with functions and loops) are easier to edit and manipulate. 

In \cite{tian2018learning}, a DSL (Domain Specific Language) is introduced for representing 3D shapes. The language supports basic primitives, the \textbf{union} (or \textbf{join}) operator and \textbf{for} loops. The introduction of \textbf{for} loops allows to capture efficiently repetitive structures, in a way similar to \cite{Nandi2020}. An interesting part of the approach is the use of self-supervised learning: A program representing a 3D shape is generated by a neural program generator, a neural program executor generates a 3D shape from the program.

\begin{figure}[!h!t!bp]
\centering
\includegraphics[width=0.95\textwidth]{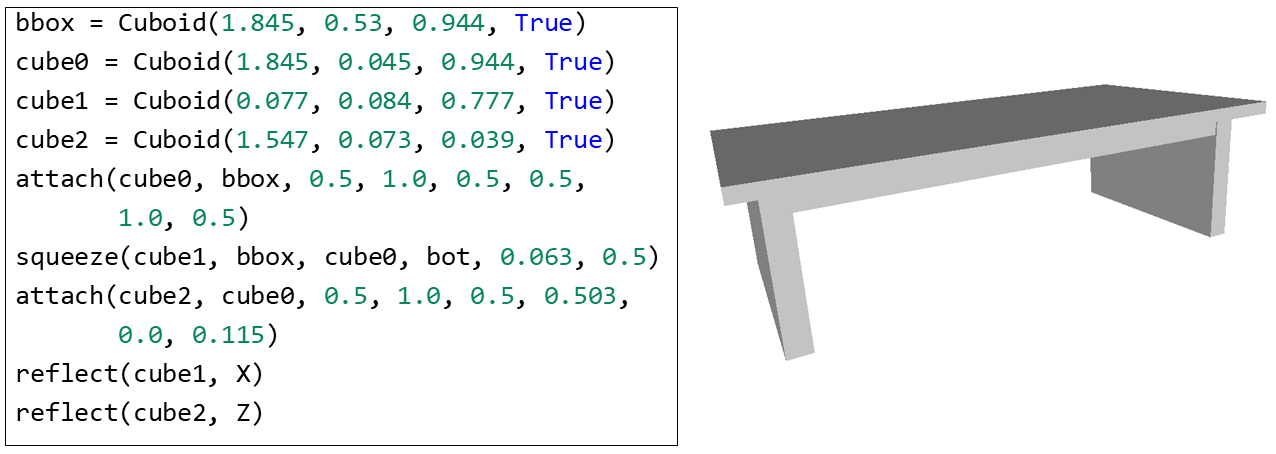}
\caption{An example of DSL from \cite{Jones2020ShapeAssembly,Jones2021ShapeMod} (left) and the corresponding 3D model (right).}\label{fig:dsl_example}
\end{figure}

The idea to decompose a shape into sub-parts is actually very similar to the decomposition of a computer program into smaller units such as functions. Techniques from deep generative models recover an assembly structure describing how parts of a shape form a whole \cite{li2020learning,Jones2020ShapeAssembly}. These approaches typically rely on training an autoencoder. The encoder generates a latent code for a given point-cloud or triangle mesh, and the decoder generates from the latent code a model expressed in a simplified domain specific language with a limited set of geometric primitives. Extracting repeating patterns can be done by introducing macros in the language and learning these macros across large data sets to help procedural modeling \cite{Jones2021ShapeMod}. For a recent survey on deep generative modeling, we refer the reader to \cite{Chaudhuri2019}. An example of DSL used in \cite{Jones2020ShapeAssembly,Jones2021ShapeMod} for the purpose of learning generative models is shown in Fig.\,\ref{fig:dsl_example}, with the corresponding geometric model shown on the right. The learning is done on collections of relatively simple shapes, and so the corresponding language can be kept relatively simple as well (cuboid primitives, with affine transformations and symmetrization operations). 

Building models by assembling parts is a common idea in computer aided design. The method described in \cite{willis2021joinable} and the related AutoMate system \cite{jones2021automate} learn to assemble parts together to form joints. The target models are expressed as B-rep models. 

Typical models, however, are often designed as a series of sketch, extrusion and Boolean operations. 
A simple language with sketch and extrude operations is described in \cite{Willis2021}. The goal is to be able to learn CAD programs in a concise programmatic form. In this work, a sequential CAD program is viewed as a Markov decision process that can be learned. This work also introduces a data set of human design sequences expressed in the language. This data set can be used for training models. 

In order to learn to generate a series of sketch, extrusion and Boolean operations corresponding to a given B-rep, Xu et al. propose to use a new geometric representation called a zone graph \cite{xu2021}. A zone corresponds to a set of solid regions derived from the B-rep faces. A graph is built with zones as nodes, and edges between nodes with geometric adjacency. Generating a model corresponds to a problem search in the space of extrusions permitted by a given zone graph. 
A related approach is DeepCAD \cite{wu2021deepcad}, where CAD models are generated as a sequence of operations for defining 2D sketches, computing their extrusion and combining them by Boolean operations. The CAD generative networks are based on \emph{transformers} \cite{vaswani2017attention}. The conversion of a point-cloud to a CAD model is not directly addressed in this work, but is tackled in \cite{uy2021point2cyl}. Given an input 3D point-cloud, a set of extrusion cylinders is generated, where an extrusion cylinder consists of a 2D sketch with an extrusion operation. 

\subsection{Challenges}
Moving forward, one of the main challenges is to extend the target representation from the CSG representation, where a solid is built by combining primitives with Boolean operations, into higher-level representations involving higher-level primitives, such as the ones obtained from sketches and extrusion, as well as operations such as blend, chamfer, bevel, linear-extrusion or sweeps. We discussed some of the recent works going in this direction in the previous section. 

CAD models can be understood as programs written in a DSL \cite{Ganin2021computeraided}. The problem of recovering a CAD model from unstructured data can then be seen as an instance of multi-modal learning, a topic with an increasing interest in artificial intelligence \cite{akkus2023multimodal}.

\subsection{Concluding remarks}
We have surveyed in this document existing methods for recovering CSG representations from unstructured data. We have covered techniques for solving related problems such as segmentation and fitting of primitives to data. We started with techniques from CAD and solid modeling for converting polyhedron and B-rep to CSG. We have looked at techniques from different domains such as program synthesis, evolutionary methods (genetic algorithms, genetic programming), or deep learning based methods. Finally, we have discussed some current research directions, such as the recovery of higher-level representations, and techniques for the generation of computer programs representing solid models, as well as some of the potential existing challenges.

\bibliographystyle{alpha}
\bibliography{refs}

\end{document}